\documentclass{kluwer}    % Specifies the document style.
\usepackage[]{graphicx}
\usepackage[]{epsfig}
\newdisplay{guess}{Conjecture}

\begin{document}
\begin{article}
\begin{opening}
\title{Gas Phase Processes Affecting Galactic Evolution}
\author{Bruce G. \surname{Elmegreen}}
\runningauthor{Bruce G. Elmegreen} \runningtitle{Gas Phase
Processes Affecting Galactic Evolution} \institute{IBM T. J.
Watson Research Center, P.O. Box 218, Yorktown Heights, NY 10598,
USA, bge@watson.ibm.com}

\begin{abstract}
Gas processes affecting star formation are reviewed with an emphasis
on gravitational and magnetic instabilities as a source of turbulence.
Gravitational instabilities are pervasive in a multi-phase medium,
even for sub-threshold column densities, suggesting that only an ISM
with a pure-warm phase can stop star formation.  The instabilities
generate turbulence, and this turbulence influences the structure and
timing of star formation through its effect on the gas distribution
and density.  The final trigger for star formation is usually direct
compression by another star or cluster.  The star formation rate is
apparently independent of the detailed mechanisms for star formation,
and determined primarily by the total mass of gas in a dense form. If
the density distribution function is a log-normal, as suggested by
turbulence simulations, then this dense gas mass can be calculated and
the star formation rate determined from first principles. The results
suggest that only $10^{-4}$ of the ISM mass actively participates in
the star formation process and that this fraction does so because its
density is larger than $10^5$ cm$^{-3}$, at which point several
key processes affecting dynamical equilibrium begin to break down. 
\end{abstract}

\end{opening}

\section{Introduction}

Gas processes in the interstellar medium (ISM) are varied and
complex. This review is limited to those most closely involved
with precursors to star formation. Other talks at this conference
cover the high energy phase and the dispersal of gas after star
formation.  Some ideas expressed here are considered in more
detail in Elmegreen (2002).

At the beginning of star formation is cloud formation, but stars
are also triggered in pre-existing clouds by processes unrelated
to cloud formation (e.g., by supernovae), and many clouds are
formed that do not produce new stars (e.g., diffuse clouds). Thus
star formation is distinct from cloud formation.

Figure 1 shows a diagram of the flow of energy into ISM structure,
starting with sources dominated by young stars, gaseous
self-gravity, and magnetism (which derives its energy from
galactic rotation via the dynamo).  The stellar sources tend to
produce expanding regions and cosmic rays, turning their energy
into radiation behind shock fronts and turbulent motions that also
decay into radiation.  Gravity produces contracting regions by
swing amplified instabilities and collapse along spiral arms. This
contraction releases more gravitational energy as the density
increases, and again much of this energy goes into turbulence and
ultimately radiation. The shells produced by stellar pressures and
the turbulence produced in these shells and by various
instabilities makes the observed cloudy structure of the ISM.
Other stellar pressures, along with continued self-gravity and
magnetic forces, then modify these clouds and eventually
produce individual and binary stars on very small scales.
\begin{figure}
\centerline{\includegraphics[width=28pc]{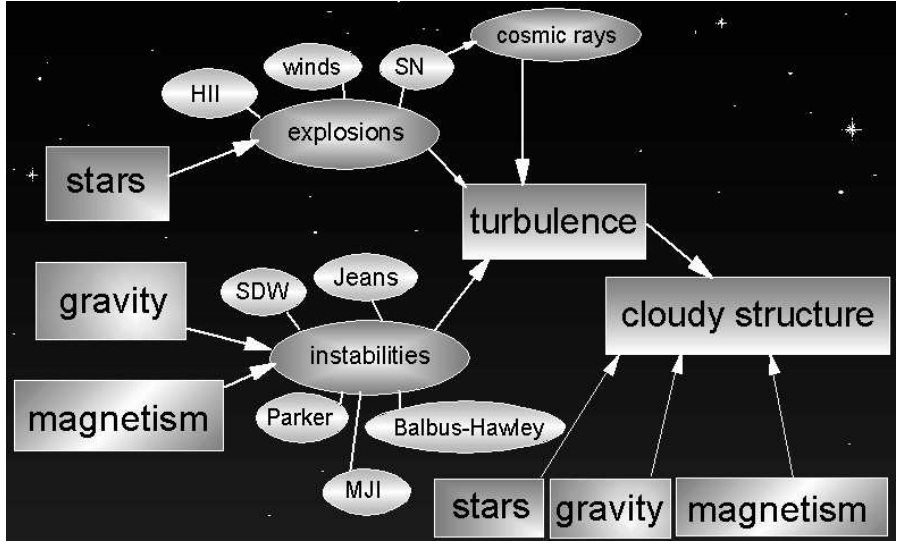}}
\caption{Schematic diagram showing paths from the main energy
sources, which are self-gravity, magnetic fields, and stars, to
the formation of cloudy structure, going through intermediate
steps of explosions, instabilities, and turbulence.  The cloudy
structure that is formed by these processes is modified further by
stars, gravity and magnetic fields.}
\end{figure}

\section{Gravitational Instabilities}

Gravitational instabilities have two characteristic lengths. One
balances pressure and self-gravity and is the Jeans length,
$2c^2/G\Sigma$, for velocity dispersion $c$ and mass column
density $\Sigma$. The other balances the Coriolis force and
self-gravity and is the Toomre (1964) length, $2\pi G \Sigma /
\kappa^2$, for epicyclic frequency $\kappa$. The characteristic
mass of the Jeans length is $c^4/G\Sigma^2$, which is $\sim10^7$
M$_\odot$ for typical stellar disks.  This is usually the largest
mass of globular clouds in galaxies (Elmegreen \& Elmegreen 1983;
Grabelsky et al. 1987; Rand 1993a,b).

The Toomre length enters into the separation between spiral arms.
Galaxies with relatively large disk-to-halo mass ratios have large
Toomre lengths and few spiral arms in a grand-design pattern.
Galaxies with relatively low-mass disks tend to have short
flocculent spiral arms (Mark 1976; Elmegreen \& Thomassen 1993;
Athanassoula, Bosma, \& Papaioannou 1987; Fuchs 2003).

Gravitational instabilities in galaxy disks involve a competition
between three forces: pressure, Coriolis, and self-gravity. The
fastest growing wavelength tends to be determined by the Jeans
length alone.  If the Coriolis force exceeds the gravitational
force at this fastest growing wavelength, then there can be no
instabilities at any wavelength. This is the threshold condition
written as $Q=c\kappa/\pi G \Sigma >1$ or $\Sigma<\Sigma_{\rm
crit}=c\kappa/\pi G$ (Safronov 1960; Toomre 1964).  Kennicutt
(1989) and Martin \& Kennicutt (2001) noted how star formation
tends to end in the outer parts of galaxy disks where this
stability threshold is first satisfied.

A similar threshold behavior arises from other processes too.
Giant shells can collapse into self-gravitating clouds that form
stars when $\Sigma$ is large compared to the same threshold,
$\Sigma_{\rm crit}$ (Elmegreen, Palou\v s, \& Ehlerov\'a 2002).
This is because Coriolis forces resist self-gravity during the
expansion of shells, causing them to twist and stall, and because
the characteristic size of a shell at the time of its internal
instability is about the Jeans length (divided by $2\pi$).

Turbulence should also show a threshold behavior considering that
the ratio $\Sigma/\Sigma_{\rm crit}$ equals $R_{\rm epi}/H$ for
epicyclic radius $R_{\rm epi}$ and disk scale height $H$ (ignoring
stars).  When the gas is sub-threshold, turbulent eddies swirl
around before they accumulate and compress enough gas to make
clouds that are gravitationally bound.

The threshold $\Sigma_{\rm crit}$ can vary by a factor of 2 around
the fiducial value of $c\kappa/\pi G$, depending on the details of
the situation.  Magnetic fields in the azimuthal direction
increase $\Sigma_{\rm crit}$ when shear is high (Gammie 1996), as
does the non-zero thickness of the disk (Romeo 1992).  Magnetic
fields reduce $\Sigma_{\rm crit}$ and make the instability more
favorable when shear is low because they remove angular momentum
from the growing condensation, offsetting the Coriolis force
(Elmegreen 1987; Kim \& Ostriker 2001). Viscosity does the same
thing for the same reason (Elmegreen 1995; Gammie 1996).  A soft
equation of state makes instabilities easier too by replacing the
velocity dispersion $c$ with the product $\gamma c$ for ratio of
specific heats $\gamma$, which can be less than 1 (corresponding
to the common observation that denser regions are cooler).
Combined stars and gas reduce $\Sigma_{\rm crit}$ because stars
support the self-gravity of the gas (Jog \& Solomon 1984; Orlova,
Korchagin, \& Theis 2002).  The Parker instability aids the
gravitational instability also by adding compressive forces along
the upward-bent magnetic field lines (Chou et al.
2000; Kim, Ryu, \& Jones 2001; Franco et al. 2002). Spiral arm
compression increases $\Sigma$ more than $\Sigma_{\rm crit}$
(which varies as $\kappa\propto\Sigma^{1/2}$ -- Balbus 1988).  A
cool sub-population of clouds can also favor instabilities, which
can act primarily in that phase (Ortega, Volkov, \& Monte-Lima
2001).

\subsection{Gravitational Instabilities in a Multi-Phase Medium}

The gravitational instability changes significantly in a
multi-phase medium. The gas can generate structure by
instabilities in the cool phase even when the bulk ISM is stable
for total rms speed $c$ (Wada \& Norman 2001).  In a diffuse cloud
population with a thermal dispersion of $\sim1$ km s$^{-1}$, the
value of $\Sigma/\Sigma_{\rm crit}$ is larger than for the bulk
medium by a factor of 10 and the unstable mass is smaller by a
factor of $10^4$.  In a turbulent medium where the dispersion
varies as the square root of length, the Jeans length can be any
length, provided the largest turbulent scale is unstable. As a
result, an ISM with a cool phase of gas available in equilibrium
should be unstable as long as $\Sigma>c_{\rm low} \kappa/\pi G$
for $c_{\rm low}$ equal to the rms dispersion in the cool phase.
This criterion is usually satisfied easily. {\it Absolute
stability therefore requires an ISM that has no cool phase}
(Elmegreen \& Parravano 1994).

The possibility of continued small scale instabilities in a
globally stable ISM raises the question of why
$\Sigma\sim\Sigma_{\rm crit}$ generally in main galaxy disks. The
usual explanation is that star formation regulates the bulk
velocity dispersion, $c$, and therefore regulates $\Sigma_{\rm
crit}$ so that the ISM is in a state of marginal stability.  When
$c$ gets too low, the instability condition is satisfied, star
formation gets more active, and $c$ increases (Goldreich \& Lynden
Bell 1965).  But if cloud-forming instabilities still operate in
the cool phase when the bulk ISM is stable ($\Sigma<\Sigma_{\rm
crit}$), then such a star formation thermostat cannot be very
effective.  More likely, turbulence generated by gravitational
instabilities keeps $\Sigma\sim\Sigma_{\rm crit}$, independent of
star formation (Fuchs \& von Linden 1998; Bertin \& Lodato 2001;
Combes 2001). In this case, there would be {\it no self-regulation
of star formation} (i.e., by other star formation), but only a
regulation of turbulence using rotation and the binding energy of
the ISM.

\subsection{An Absolute Minimum Column Density for Star Formation}
\label{sect:min}

There is growing evidence for an absolute stability condition that
corresponds to a minimum column density of around $\Sigma_{\rm
min} \sim7$ M$_\odot$ pc$^{-2}$ to get a minimum pressure that
makes a cool phase (Elmegreen 2002). Such a minimum was first
observed in irregular galaxies (Skillman 1987) and is still
evident as a threshold for star formation there (Taylor et al.
1994; Meurer et al. 1996; van Zee et al. 1997; Hunter et al. 1998;
Hunter, Elmegreen, \& van Woerden 2001; Young \& Lo 1996, 1997ab).

The outer parts of spiral galaxies tend to become pure warm where
star formation stops (Braun 1997). The faint continuation of star
formation in some galaxies beyond the $\Sigma/\Sigma_{\rm crit}<1$
threshold (Ferguson et al. 1998; LeLi\`evre \& Roy 2000) may be
the result of spiral arm compression, which can trigger either
gravitational instabilities (Kim \& Ostriker 2002) or an ISM phase
change to a cool state (Shu et al. 1972).

Several examples give some indication of the relative importance
of $\Sigma_{\rm crit}$ and $\Sigma_{\rm min}$.  O'Neil, Bothun, \&
Schombert (2000) and O'Neil, Verheijen, \& McGaugh (2000) studied
low surface-brightness galaxies with $\Sigma>\Sigma_{\rm crit}$
but not much star formation, thereby violating the usual
instability condition. SBS 0335-052 (Pustilnik et al. 2001) is
another example: the surface density is below the minimum
threshold $\Sigma_{\rm min}$ and there is not much star formation,
but the rotation speed and $\kappa$ are so low that
$\Sigma>>\Sigma_{\rm crit}$. So here again, the usual condition
does not work.  The inner parts of M33 and NGC 2403 have
$\Sigma<\Sigma_{\rm crit}$ and normal star formation (Martin \&
Kennicutt 2001), violating the usual condition again, but these
regions still exceed the absolute minimum. Similarly, the nuclear
region of the S0/E7 galaxy NGC 4550 has $\Sigma<\Sigma_{\rm crit}$
with star formation still present (Wiklind \& Henkel 2001).

The $\Sigma>\Sigma_{\rm crit}$ criterion for star formation works
most of the time in many types of galaxies, but the
$\Sigma>\Sigma_{\rm min}$ condition also works and usually both
conditions are satisfied. The few odd regions where only one or
the other threshold shows up suggest that $\sim7$ M$_\odot$ is an
absolute minimum for star formation (depending on the radiation
field) regardless of $\Sigma/\Sigma_{\rm crit}$.  This result is
consistent with the conclusion of the previous section, that
$\Sigma_{\rm crit}$ is not directly related to star formation but
more to the regulation of turbulence and cloud formation.

A different value for $\Sigma_{\rm crit}$ based on the rate of
shear rather than the epicyclic rate was recently discussed by
Pandey \& van de Bruck (1999).  The point of this threshold is to
suggest that instabilities always operate regardless of the Toomre
condition (angular momentum is removed by magnetic tension and
viscosity) and so the formation of clouds depends on the product
of the local dynamical rate and the shear time (Elmegreen 1993).
This product is approximately $\Sigma/\Sigma_{\rm crit,A}$ where
$\Sigma_{\rm crit,A}\sim cA/\pi G$ for Oort parameter
$A=-0.5Rd\Omega/dR$ in a galaxy with angular rotation rate
$\Omega$.

\subsection{Gravitational Instabilities and Accretion}

The gravity from spiral arms generates a torque on the gas and
stars that leads to an angular momentum flux and mass motions
(Larson 1984; Lin \& Pringle 1987). The angular momentum flux is
approximately $F=\pi^3G^2r^2 \Sigma^3 /\Omega^2$. Setting this
equal to $3\pi\Sigma r^2\Omega\nu$ gives an effective viscous
coefficient $\nu$ (Takeda \& Ida 2001).   The accretion time over
a distance $D$ then becomes $D^2/\nu$, which is
\begin{equation}
t_{\rm accretion}\sim {{3D^2\Omega^3}\over{\pi^2G^2\Sigma^2}}\sim
{{1.5D^2Q^2\Omega}\over{c^2}},\end{equation} where the latter
expression is for a flat rotation curve. In this case, and with
$Q\sim1$ for most galaxy disks, the accretion speed is
\begin{equation}
v_{\rm accretion}\sim 0.6 c \left(c/D\Omega\right),
\end{equation}
that is, the rms turbulent speed times the ratio of this turbulent
speed to the shear speed over the distance $D$.

The accretion time may be evaluated for a few interesting cases.
It equals about 500 My over the disk Jeans length (several kpc)
for stars in the main disks of spiral galaxies.  The resulting
accretion liberates gravitational energy from the rotation of the
galaxy and heats the stars. As a result, spiral instabilities soon
stop unless cool stars are added (Sellwood \& Carlberg 1984). The
accretion time is about 1 Gy over one kpc for the gaseous parts of
galaxy disks, considering a column density corresponding to
$A_V=0.5$ mag.  This is significantly shorter than the Hubble
time, suggesting that accretion can change the surface density
profiles of disks (Lin \& Pringle 1987; Yoshii \& Sommer-Larsen
1989; Saio \& Yoshii 1990; Gnedin, Goodman, \& Frei 1995; Ferguson
\& Clarke 2001) and amplify a metallicity gradient. The accretion
time is $\sim1$ Gy over 100 pc in nuclear regions with faint dust
disks ($A_{\rm V}\sim1$ mag; corresponding to $Q>>1$), but can be
very short, $\sim30$ My over 100 pc in nuclear starburst regions
where $Q\sim1$ (because of the very high $\Sigma$).

\subsection{Gravitational Instabilities and Turbulence}

The energy liberated by ISM collapse into spiral arms drives
turbulence on kpc scales.  The energy liberated by cloud collapse
drives turbulence on smaller scales. Thus the entire cascade of
turbulent structures can be driven by self-gravity.  Stellar
pressures contribute to turbulence more locally.  This means that
$\Sigma\sim\Sigma_{\rm crit}$ could be controlled by spiral
instabilities, not star formation feed-back, as discussed above.
Numerical simulations and other theoretical work on this source of
turbulence may be found in Fuchs \& von Linden (1998), Semelin \&
Combes (2000), Bertin \& Lodato (2001), Wada \& Norman (2001),
Huber \& Pfenniger (2001, 2002), Vollmer \& Beckert (2002), and
Chavanis (2002).

Crosthwaite, Turner, \& Ho (2000) pointed out that some of the
holes in the interstellar medium of IC 342 can be generated by
gravitational instabilities.  Stellar pressures are not always
necessary to make holes.  Wada, Spaans, \& Kim (2000) also got ISM
holes in a simulation with turbulence and self-gravity.

Gravitational instabilities are energetically important as a
source of turbulence for the ISM. The power density is
approximately the ISM energy density multiplied by the
gravitational instability growth rate, $\pi G\Sigma/c$. This
amounts to $2\times10^{-27}$ erg cm$^{-3}$ s$^{-1}$.  The amount
is large because the growth time is short, $c/\pi G\Sigma\sim30$
My. This is also about the energy dissipation time. Vollmer \&
Beckert  (2002) show that the energy flux to small scales from
turbulence equals the energy dissipation by self-gravitational
instabilities in the disk.

The energy from spiral chaos, as Toomre \& Kalnajs (1991) called
this process, is comparable to that from supernovae at a rate of
one per 100 years in a whole galaxy, assuming a 1\% efficiency for
converting the explosion energy into ISM motions.

\section{The Parker and Balbus-Hawley Instabilities}

Magnetic buoyancy and cosmic ray pressure push magnetic field
lines out of the galactic plane, and then the gas trickles down
into the valleys, forming clouds with low density contrast (Parker
1966). This is an instability because the more the field lines
buckle, the greater the streaming speed and collection of gas into
the valleys. The unstable growth time is about the propagation
time of an Alfv\'en wave over one scale height.

This is not a particularly good cloud formation mechanism by
itself because the flows remain in pressure equilibrium and the
maximum density contrast is essentially equal to the ratio of the
total pressure to the pure gas pressure, which is only a factor of
3. However, in combination with self-gravity, this mechanism can
help to form giant cloud complexes along spiral arms. In this
sense, both the gravitational instability and the Parker
instability work together. They have comparable time scales and
length scales along the mean field direction. Recent simulations
of the Parker instability are in Chou et al. (2000), Kim, Ryu, \&
Jones (2001) and Franco et al. (2002).

The Balbus \& Hawley (1991) instability works for either azimuthal
or vertical magnetic fields, with the former applicable to
galactic disks. The field couples regions at different radii and
transfers angular momentum directly from the inner disk to the
outer disk, which is moving at a different angular speed. The
outer disk gains angular momentum and goes out even further. This
is not a cloud-forming instability but it can generate turbulence
at the Alfven speed. The growth time is about an orbit time, and
the energy input rate is $\sim0.6(B^2/8\pi)\Omega$ for galactic
rotation rate $\Omega$ and magnetic field strength $B$ (Sellwood
\& Balbus 1999).  Sellwood \& Balbus suggest that this instability
could drive turbulence in the outer parts of galactic disks where
the supernova rate is very low and there are no other energy
sources.  If the outer disks are a nearly pure-warm phase of HI,
however (Sect. \ref{sect:min}), then the modest velocity
dispersions observed there can be thermal, in which case no source
of turbulence is needed.

\section{Clouds from Shells or Turbulence?}

HI maps of Ho II (Puche et al. 1992),  IC 2574 (Walter \& Brinks
1999; Steward \& Walter 2000), the LMC (Kim, et al. 1999;
Yamaguchi et al. 2001a), and other small galaxies show shell-like
structures covering a large fraction of the interstellar volume.
Shear is low in all of these cases, as is the ambient pressure, so
common supernova explosions and HII region expansions can make
these shells and inflate them to large sizes without severe
distortion.  The expansions may even continue for so long that the
stars which initially made them disperse.

The power spectrum of the HI emission from the LMC is a power law
with a slope characteristic of turbulence (Elmegreen, Kim, \&
Staveley-Smith 2001). This seems odd if the HI structure is
entirely the result of expanding shells caused by star formation.
An alternative model is that the structure comes from turbulence
that is indirectly generated by young stars (Wada, Spaans, \& Kim
2000).  Such turbulence can still make hot shells, but there will
not be a one-to-one correspondence between these shells and the OB
associations.

Galaxies with higher shear tend to show spiral arms instead of
shells. These arms may begin as shells and then get swept back
into spiral shapes, or they may have other origins.  The faint
dust spirals in the nuclei of two early type galaxies, NGC 4736
and NGC 4450, have the same power-law power spectra as the HI
emission from the LMC (Elmegreen, Elmegreen \& Eberwein 2002). The
spirals are probably turbulent in origin, but they do not appear
to be connected with star formation, which has a relatively small
rate in these high-Q regions.

These observations suggest there are two types of global ISM
structures: shells that are made directly or indirectly by the
energy of star formation, and shells or spirals that are made by
turbulence originating with instabilities.  The first type tends
to show up in regions of low shear.  This special position could
mean that shear alone determines the morphology of clouds. It
could also mean that the instabilities depend on shear and vanish
when the shear rate is low, leaving only star-formation shells.

\section{Correlated Structures in Young Star Fields}

Interstellar turbulence from gravitational and other instabilities
plus star formation and other pressure sources makes
autocorrelated, multi-fractal structure in the gas. This structure
may be interpreted as clouds in an intercloud medium, but the
cloudy description is often too simple and can lead to selection
effects (Scalo 1990).  The same autocorrelated structures appear
in young star fields because star formation follows the gas
(Heydari-Malayeri et al. 2001; Pietrzynski et al. 2001; Elmegreen
\& Elmegreen 2001; Zang, Fall, \& Whitworth 2001). The resulting
stellar patterns can lead to selection effects.  Most likely,
flocculent spiral arms, star complexes, OB associations, and OB
subgroups are equivalent parts of a continuum of structures that
extend from the galactic scale height down to the sub-parsec
region where dense embedded clusters form (Elmegreen et al. 2000;
Elmegreen, Elmegreen \& Leitner 2003).

Because of the turbulent origin for much of the ISM structure, the
dynamical time for motions varies approximately as the square root
of the region size.  A similar scaling occurs for star formation:
the duration of star formation in a region increases approximately
as the square root of the size (Efremov \& Elmegreen 1998), always
being a few dynamical times (Ballesteros-Parades et al. 1999;
Elmegreen 2000).

Stars and clusters form in the densest part of the ISM fractal
(Heyer, Snell, \& Carpenter 1997) where the gas is molecular
because of dust and self-shielding, cold because it is molecular
and cools well, and strongly self-gravitating because it is dense
and cold. The mass functions for both clusters and clouds are
power laws because the motions that make them are turbulent and
turbulence makes self-similar structures, which have power-law
size distributions (Elmegreen 2002).

\section{Triggered Star Formation}

Most local clusters look triggered by adjacent HII regions.
Yamaguchi et al. (1999, 2001b) estimate that 10\%-50\% of inner
Milky Way star formation, and the same fraction of LMC star
formation, is triggered by expanding HII regions.  What is the
connection, then, between star formation and turbulence?

Instabilities drive turbulence globally and stellar pressures
drive turbulence locally. The instabilities and turbulence
together make clouds with a wide range of scales.  No Jeans mass
is evident except for the ``beads'' of star formation in stellar
spiral arms.  Stellar pressures make shells and shape the existing
clouds into comets, triggering star formation. The time scale for
instabilities is about equal to the crossing time from turbulence,
and this is about equal to the time scale for triggering.  Thus
one group of processes (instabilities and shell-formation) makes
clouds, while another group of processes (pressurized-triggering)
often makes stars in these clouds.  The time scale is about the
same for each, always comparable to the dynamical time.

\section{The Star Formation Rate from First Principles}

Stars form only at high density yet the star formation rate scales
with the average density, $\rho$,
\begin{equation}
{\rm SFR(mass/vol/time)}\sim\epsilon\rho(G\rho)^{1/2}
\end{equation}
for efficiency $\epsilon$. In high density cores, the star
formation rate should be
\begin{equation}
{\rm SFR(mass/vol/time)}\sim\epsilon_c\rho_c(G\rho_c)^{1/2}\sim
\epsilon_c \rho_c\omega_c
\end{equation}
for core efficiency $\epsilon_c$, density $\rho_c$, and rate
$\omega_c$.

For a threshold core density $\rho_c\sim10^5$ cm$^{-3}$ and
typical $\epsilon_c\sim0.1$-0.5, the core star formation rate is
constant. At $\rho_c\sim10^5$ cm$^{-3}$, big grains stop gyrating
(Kamaya \& Nishi 2000), molecules freeze onto grains (Bergin et
al. 2001), the ionization fraction begins to drop (Caselli et al.
2002), and turbulence becomes subsonic (Goodman et al. 1998).

With these relations, the Schmidt-law implies that $\rho_c/\rho$
is constant and that $\epsilon$ is proportional to the fraction of
the gas at $\rho>\rho_c$.  If the Schmidt law is not correct, but
instead stars form at constant efficiency (Rownd \& Young 1999;
Boselli et al. 2002), then $\rho_c$ is constant and $\epsilon$ is
still proportional to the fraction of the gas at greater density.

Wada \& Norman (2001) found a log-normal probability distribution
function (pdf) for density in their whole galaxy models. The dense
gas fraction is the integral over this distribution function above
the density threshold. Elmegreen (2002) normalized this pdf to
the local density, and then normalized the Kennicutt (1998)
Schmidt law to the local density.  After these normalizations, the
fraction of all the interstellar gas that is forming stars at the
local dynamical rate, $(G\rho_c)^{1/2}$, turns out to equal the
fraction of the ISM pdf with a density larger than $10^5\rho_{\rm
ave}$. This mass fraction is $10^{-4}$ of the ISM. If we multiply
this by the core-to-average rate ratio,
$\left(\rho_c/\rho\right)^{1/2}\sim300$, then we get the average
efficiency of star formation over the average dynamical time in
the ISM; this average efficiency is a reasonable $\sim3$\%.

The point of this exercise is to show that the Kennicutt-Schmidt
law on a galactic scale can arise from numerous events of local
star formation, each on the scale of an individual cloud core, if
the high formation rate in each core is averaged out over all the
ISM gas, considering only the gas that is participating in the
star formation process.   The core density that is necessary to do
this is a reasonable $10^5$ cm$^{-3}$, which is where we think
star formation begins anyway, and the fraction of the ISM that has
this density or greater is about $10^{-4}$ if the Wada \& Norman
pdf shape is correct. In each core, the efficiency of star
formation is high, but averaged over all the ISM, it is low,
several per cent.

\section{Conclusions}

Instabilities involving gravity, magnetism, and pressure lead to
spirals, accretion, clouds, and turbulence.  Stellar pressures
produce bubbles, more turbulence, and triggered star formation in
clouds that already formed. Self-gravity and turbulence combine to
structure the ISM, giving self-correlated properties for the gas
and young stars with respect to size, velocity dispersion, and
crossing time or duration of star formation. Turbulence also gives
power law mass functions for clouds and clusters.

The turbulence generated by gravitational instabilities can
maintain the ISM in a state of quasi-equilibrium where
$\Sigma\sim\Sigma_{\rm crit}$.  If small scale instabilities
continue in the cool component of the gas even when the average
rms speed is large enough to give global stability, then star
formation cannot regulate the $\Sigma/\Sigma_{\rm crit}>1$
threshold. In this case, there is no self-regulation of star
formation involving $\Sigma_{\rm crit}$ on a galactic scale. This
will be true even if young stellar pressures agitate the ISM
locally. They can blow the gas out into the halo and stop star
formation locally, but young stars probably cannot fine-tune or
moderate their own formation rate so that it stays near the
historical or Hubble-type average. Young stars commonly trigger
other stars anyway, so the feedback they produce should
de-stabilize, not stabilize, the star formation rate,
unless the entire local ISM is removed.

The star formation rate depends on the mass fraction in dense gas.
Turbulence may determine this mass fraction, independent of the
sources for the turbulence. The global SFR is then independent of
the detailed triggering mechanisms. Then again there would be no
self-regulation of star formation, only a star formation {\it
saturated} to its maximum possible value, as determined
by the open and tenuous
geometry of the gas.  In this case, star formation can be halted
only by a dominance of the warm phase of the ISM.

\end{article}
\end{document}